\documentclass[reprint,amsmath,amssymb,aps,prl]{revtex4-1}

\usepackage{graphicx}% Include figure files
\usepackage{dcolumn}% Align table columns on decimal point
\usepackage{bm}% bold math
\usepackage{hyperref}% add hypertext capabilities

\begin{document}

\title {
    Intrinsic Band Gap and Electrically Tunable Flat Bands\\
    in Twisted Double Bilayer Graphene
}
\author {
  Young Woo Choi and Hyoung Joon Choi
}
\email {
  h.j.choi@yonsei.ac.kr
}
\affiliation {
  Department of Physics, Yonsei University, Seoul 03722, Korea
}

\date{\today}

\begin{abstract}
    We present atomistic calculations on structural and electronic
    properties of twisted double bilayer graphene (TDBG)
    consisting of two sets of rotationally misaligned Bernal-stacked 
    bilayer graphene.
    Obtained equilibrium atomic structures exhibit in-plane strains and
    the modulation of the interlayer distances at the rotationally
    mismatched interface layers.
    We find that the electronic structure of TDBG can have an intrinsic band gap
    at the charge neutral point for a large range of the twist angle $\theta$.
    Near $\theta=1.25^\circ$, the intrinsic band gap disappears and TDBG 
    hosts flat bands at the Fermi level
    that are energetically well separated from higher and lower energy bands.
    We also show that the flat bands are easily tunable by applying 
    vertical electric fields, and extremely narrow bandwidths 
    less than 10~meV can be achieved for the electron-side flat bands 
    in a wide range of the twist angle. Our results serve as a theoretical 
    guide for exploring emergent correlated electron physics
    in this versatile moir\'e superlattice system.
\end{abstract}

\maketitle

%%%%%%%%%%%%%
%%% Intro %%%
%%%%%%%%%%%%%
Moir\'e superlattices in two-dimensional van der Waals heterostructures have 
become a new experimental platform for studying correlated electron physics,
featuring various degrees of freedom that can be externally controlled.
In the case of twisted bilayer graphene (TBG),
bandwidths of Dirac electrons in graphene are tunable
through the fine control of the twist 
angle~\cite{Magaud:2010,Barticevic:2010,MacDonald:2011}
and flat bands emerge near so-called magic angles.
Recent series of experiments have shown that such magic-angle twisted bilayer 
graphene (MA-TBG) hosts correlated insulating states
at the half fillings of the flat bands and superconductivity upon doping
additional electrons or holes near the insulating 
phases~\cite{Cao:2018a,Cao:2018b}.
These observations have established that the magic-angle twisted bilayer
graphene is a promising platform for exploring correlated electron physics.

Furthermore, MA-TBG has a great advantage in terms of tunability because 
it is possible to explore the correlated phases at different doping levels 
with a single device, owing to the electrical controllability of the carrier 
concentration. In addition, the necessity of the fine control of the twist 
angle can be further relaxed by application of hydrostatic pressure to TBG 
samples. It has been theoretically suggested that the interlayer coupling in 
TBG can be enhanced by applying pressure, inducing flat bands in a large range 
of the twist angle rather than at certain discrete magic 
angles~\cite{Kaxiras:2018a,Chittari:2019}.
Furthermore, a recent experimental work has shown that, by applying pressure,
the correlated insulating states and superconductivity are induced at a
twist angle larger than the magic angle, where the correlated phases are absent
without the pressure~\cite{Dean:2019}.

So far limited attention has been paid to the layer-number degree of freedom. 
Since the number of layers is an important degree of freedom that strongly 
affects physical properties of van der Waals materials, it can also provide 
an additional dimension to the flat-band physics in moir\'e superlattices.
In particular, AB-stacked bilayer graphene (AB-BLG) has a widely tunable
band gap that can be controlled by a vertical electric 
field~\cite{Ohta:2006,Neto:2007,Wang:2009}.
Therefore, one can expect that a moir\'e superlattice formed by two sets of
AB-BLG inherits such tunability of the electronic structure with an external 
bias, as in continuum models~\cite{Zhang:2019,Chebrolu:2019}.

In this work, we present an atomistic study on the electronic structure of 
twisted double bilayer graphene (TDBG) consisting of two sets of AB-stacked 
bilayer graphene with a twist. Considering both in and out of plane 
structural relaxations, we calculate the electronic structure of TDBG at small 
twist angles and discuss the similarities and differences with TBG.
We find that TDBG has an intrinsic band gap at the charge neutral point
over a large range of the twist angle, owing to the absence of the inversion 
symmetry. We also show that TDBG hosts flat bands near the Fermi level at 
small twist angles similar to TBG, but their bandwidths can be further tuned by
a moderate vertical electric field.
As a result, we predict that extremely narrow bandwidths less than 10 meV
can be achieved for a large range of the twist angle.

%%%%%%%%%%%%%%%%%%%%%%%%%%%%%%%%%%%%%%%%%
%%% Figure 1 Relaxed Atomic Structure %%%
%%%%%%%%%%%%%%%%%%%%%%%%%%%%%%%%%%%%%%%%%
\begin{figure}
    \includegraphics[width=8.5cm]{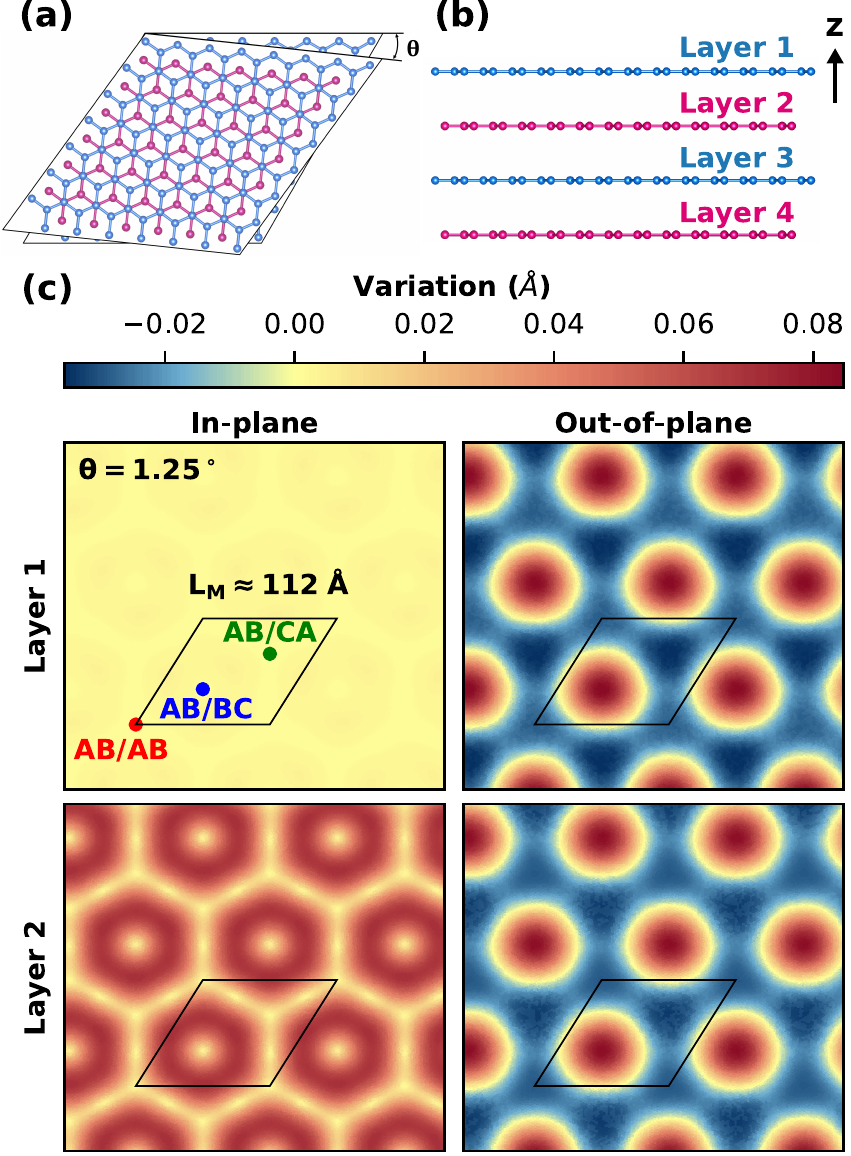}
    \caption {
      Schematic (a) top and (b) side view of the twisted double
      bilayer graphene. (c) In-plane (left) and out-of-plane (right)
      variation of atomic positions in the first (top) and second
      (bottom) graphene layers after structure relaxation. 
      The remaining two layers have similar variation patterns, 
      except that atomic positions are varied in opposite directions.
    }
    \label{fig:lattice}
\end{figure}

The commensurate moir\'e supercell of twisted double bilayer graphene
is constructed by stacking a rotated AB-stacked bilayer graphene
on the other AB-BLG, which thus consists of four graphene layers
in total [Figs.~\ref{fig:lattice}(a) and \ref{fig:lattice}(b)].
The resulting crystal structure has threefold rotation
symmetry around the $z$ axis, but lacks the inversion symmetry.
The local stacking orders of TDBG vary across the moir\'e supercell.
For instance, $AB/AB$, $AB/BC$ and $AB/CA$ stacking regions are denoted in
the upper left plot of Fig.~\ref{fig:lattice}(c).

Preserving the crystal symmetries and the supercell lattice vectors,
we determined equilibrium atomic positions of TDBG
by minimizing the total energy $U$ consisting of
the in-plane elastic energy and the interlayer van der Waals
binding energy (See Ref.~\cite{Choi:2018} for the detailed description
of the structural relaxation method).
The in-plane elastic energy is calculated under the harmonic approximation
where atomic force constants are considered up to the 4th nearest neighbors,
as obtained by fitting \textit{ab initio} phonon dispersion 
calculations~\cite{Wirtz:2004}. The interlayer van der Waals binding energy 
is calculated
using Kolmogorov-Crespi (KC) potential that depends on interlayer atomic
registry~\cite{Kolmogorov:2005}. Our method has been shown to be effective
in obtaining relaxed atomic positions of magic-angle twisted bilayer 
graphene~\cite{Choi:2018},
and can be applied straightforwardly to twisted multilayer graphene systems.

The rotational mismatch at the interface of two bilayer graphene
results in both in and out of plane variations of atomic positions.
Figure~\ref{fig:lattice}(c) shows variations in the atomic positions
of TDBG at $\theta=1.25^\circ$ after the structural relaxation.
The in-plane variation from the perfect honeycomb structure occur mostly 
for the inner two graphene layers and almost negligible in the outer two layers.
In contrast, the out-of-plane variation is present in all of the four
graphene layers. The interlayer distance between the second and third layer, 
where the rotational mismatch is present, varies across the moir\'{e} supercell and is largest (smallest) at $AB/BC$ ($AB/CA$) stacking regions.
Atoms in the first layer relax vertically following the same out-of-plane
displacement pattern of the second layer. Thus, the interlayer distance
between the first and the second layer is nearly the same as that of
AB-stacked bilayer graphene. Relaxation patterns of the third (fourth) layer 
are similar to those of the second (first) layer, except that they relax in 
the opposite direction.

%%%%%%%%%%%%%%%%%%%%%%%%%%%%%%%%%%%%%
%%% Figure 2 Electronic Structure %%%
%%%%%%%%%%%%%%%%%%%%%%%%%%%%%%%%%%%%%
\begin{figure}
    \includegraphics[width=8.5cm]{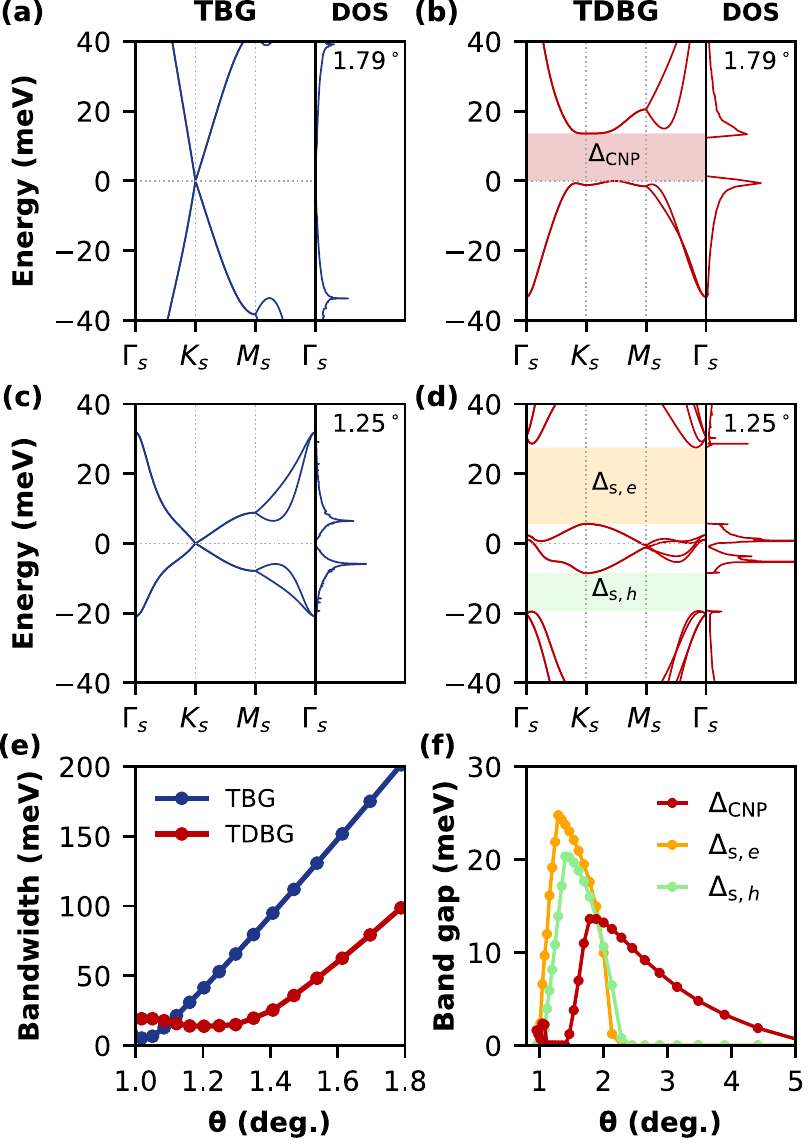}
    \caption {
    Band structures and the density of states (DOS) of
    (a,c) twisted bilayer graphene and
    (b,d) twisted double bilayer graphene with relaxed atomic structures
    at twist angles of (a,b) 1.79$^\circ$ and (c,d) 1.25$^\circ$.
    In (a-d), the absolute values of DOS are normalized to the same unit of
    states/meV/$\mathrm{nm}^{-2}$ so that their magnitudes can be directly
    compared. Plots of (a-d) in a wider energy range are given in Fig.~S1
    in the Supplemental Material~\cite{supplemental}.
    (e) The bandwidths of the flat bands in TBG and TDBG
    as a function of the twist angle $\theta$.
    (f) The band gap $\Delta_{\mathrm{CNP}}$ at the charge neutral point 
    and the band gap $\Delta_{s,e(h)}$ between the electron (hole)-side 
    flat bands and higher (lower) energy bands as functions of $\theta$ in TDBG.
    }
    \label{fig:band}
\end{figure}

With the relaxed atomic structures, we investigate the electronic structure
of TDBG and compare it with TBG using an atomistic tight-binding 
approach~\cite{Choi:2018,Moon:2012,Ando:2000}. Figures~\ref{fig:band}(a) and 
\ref{fig:band}(b) show tight-binding band structures of TBG and TDBG 
at $\theta=1.79^\circ$, respectively. At this angle,
we note two important differences between the electronic structures of
TBG and TDBG.
First, the bandwidth of TDBG is much smaller than that of TBG at the same angle.
Second, an intrinsic band gap ($\Delta_{\mathrm{CNP}}$) can open at the charge 
neutral point in TDBG, owing to the absence of the inversion symmetry.
Notably, the valence and conduction band edges become extremely flat
after the opening of $\Delta_{\mathrm{CNP}}$, and the electronic density of
states is significantly enhanced near the band edges accordingly.
The size of $\Delta_{\mathrm{CNP}}$ can be as large as 13.6~meV, 
as shown in Fig.~\ref{fig:band}(f).

When we reduce the twist angle down to $\theta=1.25^\circ$, TDBG shows nearly 
flat bands at the Fermi level while flat bands in TBG are more than two 
times wider [Figs.~\ref{fig:band}(c) and \ref{fig:band}(d)]. 
The energy gap ($\Delta_{\mathrm{CNP}}$) at the charge neutral 
point is now closed at this angle because bands overlap in some parts of 
the Brillouin zone other than near the $K_s$ point as a result of strong band 
flattening. The band splitting at the $K_s$ point still exists, meaning that 
the effect of the inversion symmetry breaking persists.
Similar to TBG~\cite{Nam:2017,Choi:2018},
the lattice relaxation in TDBG can open band gaps above
the electron-side flat bands and below the hole-side flat bands, as
denoted respectively by $\Delta_{s,e}$ and $\Delta_{s,h}$ in Fig.~\ref{fig:band}(d).
The size of $\Delta_{s,e(h)}$ at this angle is about 21.9 (10.8) meV so that
the flat bands in TDBG are well separated from the high energy bands.

In the Supplemental Material~\cite{supplemental},
Fig.~S1 shows band structures of TBG and TDBG
in a wider energy window, clearly showing $\Delta_{s,e(h)}$
in $\theta=1.79^\circ$ and $\theta=1.25^\circ$.
Furthermore, in Fig.~S2, we compare effects of the lattice relaxations
on the electronic structure. In particular, we find that the size of 
the intrinsic band gap $\Delta_{\mathrm{CNP}}$ does not depend on
the relaxation.

Bandwidths and band gaps as functions of the twist angle are summarized
in Figs.~\ref{fig:band}(e) and \ref{fig:band}(f), respectively.
The flat bands are generally narrower in TDBG than TBG for $\theta > 1.1^\circ$.
The bandwidth in TDBG reaches a minimum value near $\theta=1.25^\circ$,
which can be regarded as the first magic angle of TDBG.
The energy gap $\Delta_{\mathrm{CNP}}$ in TDBG starts to develop 
near $\theta \sim 5^\circ$ and it reaches a maximum value of 13.6 meV 
at $\theta=1.79^\circ$. Then $\Delta_{\mathrm{CNP}}$ is closed 
near $\theta=1.4^\circ$ and reopens below $\theta\sim1.1^\circ$.

%%%%%%%%%%%%%%%%%%%%%%%%%%%%%
%%% Figure 3 Bias Effects %%%
%%%%%%%%%%%%%%%%%%%%%%%%%%%%%
\begin{figure}
    \includegraphics[width=8.5cm]{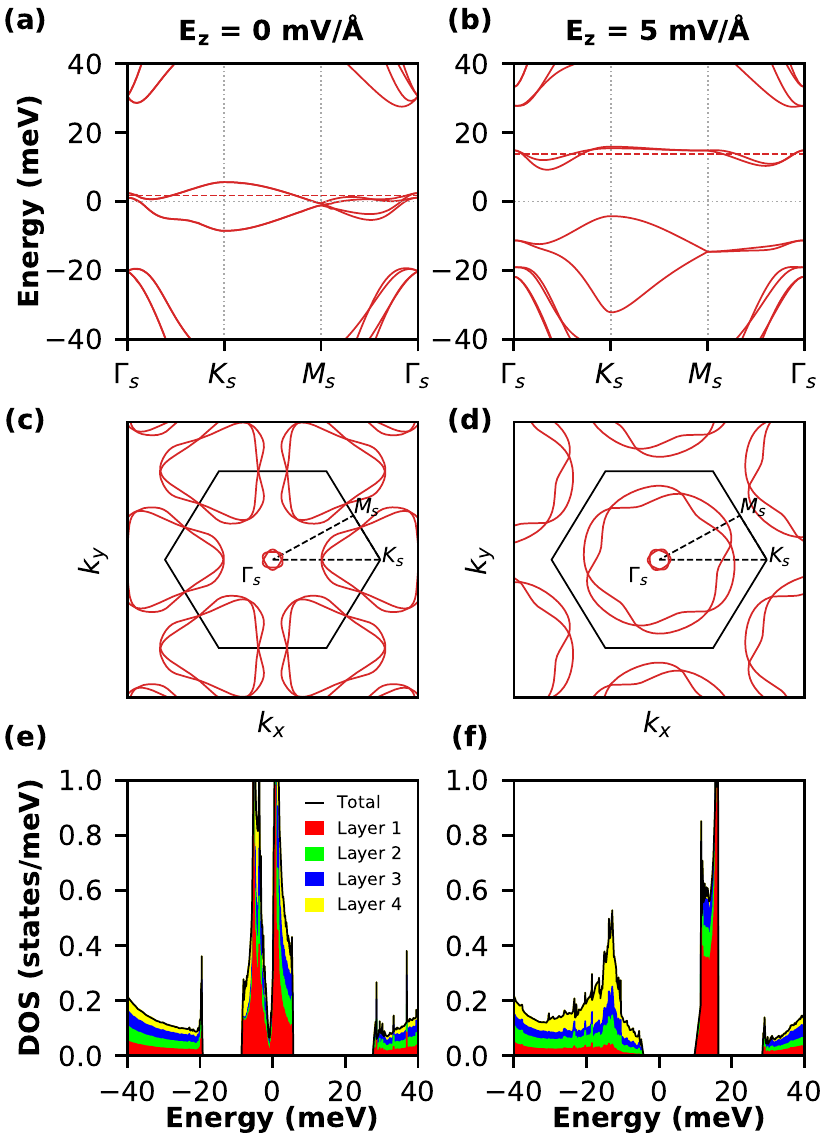}
    \caption {Electronic structures of TDBG at $\theta=1.25^\circ$. 
    (a,b) Band structures with the vertical electric field $E_z$ of 
    (a) 0~mV/{\AA} and (b) 5~mV/{\AA}.
    Horizontal dashed lines denote half-filling energies of 
    electron-side flat bands.
    (c,d) Fermi surfaces at the half-filling energies of the electron-side
    flat bands with $E_z$ of (c) 0~mV/{\AA} and (d) 5~mV/{\AA}.
    (e,f) Electronic density of states with $E_z$ of (e) 0~mV/{\AA} 
    and (f) 5~mV/{\AA}.  Contributions of the 1st, 2nd, 3rd, and 4th 
    graphene layer to the total density of states are shown in red, green, 
    blue, and yellow, respectively.
    }
    \label{fig:bias}
\end{figure}

Now we investigate how a vertical electric field modifies the electronic
structure of TDBG.
We consider a vertical electric field by adding the following 
electrostatic energy term to the tight-binding Hamiltonian:
\begin{equation}
  \Delta H = e E_z z,
\end{equation}
where $e>0$ is the elementary charge, $E_z$ is the electric field strength,
and $z$ is the $z$-coordinates of atoms.
Essentially the electric field makes on-site energy differences
in each layer.

Figures~\ref{fig:bias}(a) and \ref{fig:bias}(b) show the band structures
at $\theta=1.25^\circ$ with $E_z$ = 0 and 5 mV/\AA, respectively.
Without the electric field, all of the four flat bands are present
near the Fermi level and no band gap exists among them.
When the vertical electric field of $E_z$ = 5 mV/{\AA} is applied,
the flat bands are split into two groups and a band gap opens between them.
Notably, the electron-side flat bands are much flatter than the hole-side
flat bands, being separated from other bands.
In contrast, the electric field makes the hole-side flat bands wider
so they become overlapped in energy with the lower energy bands.

Fermi surfaces at half-filling energies of the flat bands are
of special interest because correlation effects are often maximized at
half-filling and TBG actually showed correlated insulating states 
at half-filling~\cite{Cao:2018a}.
Figures~\ref{fig:bias}(c) and \ref{fig:bias}(d) show Fermi surfaces in TDBG
at the half-filling of the electron-side flat bands
without and with the electric field, respectively.
The half-filling energy of each case is denoted by a dashed red line
in Figs.~\ref{fig:bias}(a) and \ref{fig:bias}(b).
When the external electric field is not applied,
two triangular hole pockets are located at corners of the Brillouin zone,
and additional hole pockets exist near the $\Gamma$ point
[Fig.~\ref{fig:bias}(c)].
When the electric field is applied,
Fermi surfaces become more or less circular,
and all the Fermi surfaces are centered at the $\Gamma$ point
[Fig.~\ref{fig:bias}(d)].

The electric field also induces asymmetry in the layer distributions
of the flat bands. Figure~\ref{fig:bias}(e) shows the layer-projected 
density of states without the external electric field. The flat-band 
states are rather uniformly distributed over the four layers.
However, the electric field along the $+z$ direction polarizes the flat-band 
states so that the electron (hole)-side flat bands become mostly confined 
within the first (fourth) layer. This effect is illustrated by the changes 
in the relative weights of each layer in the layer-projected density of states,
as shown in Fig.~\ref{fig:bias}(f). Along with such layer-polarization,
the density of states is significantly enhanced for the electron-side flat 
bands. We expect that this large density of states,
which are mainly localized in the first layer, would make 
correlation effects much stronger in the electron-side flat bands than 
the hole-side ones.
For comparison, we also performed the same calculations using a nonrelaxed 
structure at the same twist angle $\theta=1.25^\circ$, as shown in 
Fig.~S3 in the Supplemental Material~\cite{supplemental}, where 
the flat bands in the nonrelaxed structure also show similar electrical
tunability although they have slightly larger bandwidths and nonzero
overlaps with high energy bands.

%%%%%%%%%%%%%%%%%%%%%%%%%%%%%%%%%%%%%%%%%%%%%%
%%% Figure 4 Bandwidth vs. Electric Fields %%%
%%%%%%%%%%%%%%%%%%%%%%%%%%%%%%%%%%%%%%%%%%%%%%
\begin{figure}
    \includegraphics[width=8cm]{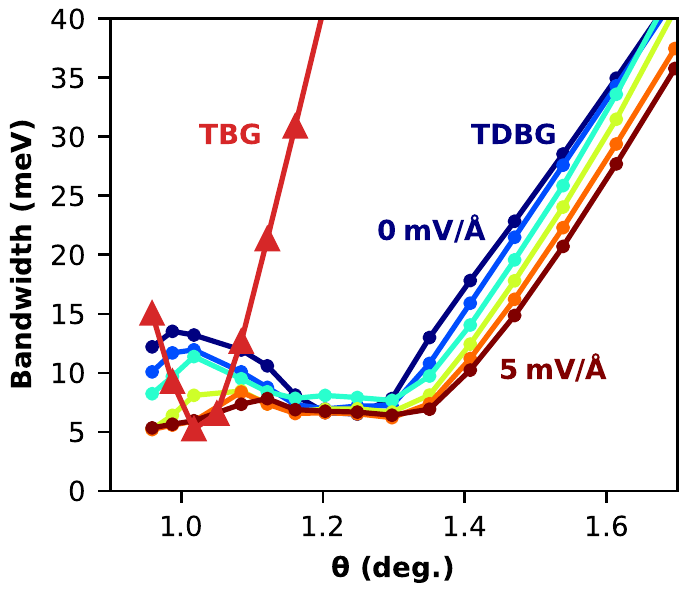}
    \caption {
    Bandwidths of flat bands in TDBG at different
    twist angles $\theta$ and vertical electric fields $E_z$.
    Here $E_z$ is increased from 0 to 5~mV/{\AA} by a step of 1~mV/{\AA}.
    For $E_z \geq 1$~mV/{\AA}, bandwidths of the electron-side flat bands
    are plotted, which are separated from the hole-side ones.
    Bandwidths of the flat bands in TBG are also plotted for comparison.
    }
    \label{fig:bandwidth}
\end{figure}

We elaborated more on the electrical tunability.
Figure~\ref{fig:bandwidth} shows the bandwidths of the flat bands
as a function of the twist angle and the strength of the vertical electric 
field. Without the electric field, the bandwidth of TDBG reaches the minimum 
value at the twist angle of $\theta\sim1.25^\circ$ and then increases again
as the twist angle is further lowered 
[Figs.~\ref{fig:band}(e) and \ref{fig:bandwidth}]. When the electric field 
is applied, the bandwidth of TDBG is continuously reduced for the most of 
twist angles (Fig.~\ref{fig:bandwidth}). For all the twist angles in the 
range of $\theta=0.96\sim1.4^\circ$, applied electric fields can make the 
bandwidth of the electron-side flat bands smaller than 10~meV, which is 
similar to the minimum bandwidth that can be obtained at the magic-angle 
twisted bilayer graphene.
This strongly suggests that correlated electron physics can emerge in TDBG
in a wide range of the twist angle rather than at certain special
magic angles.
We also calculated the electronic structure of TBG under the external electric
field and obtained that the electric-field strength considered in our present 
work has almost negligible effects in the TBG case,
which is consistent with a previous study~\cite{Moon:2014}.
Thus, the strong tunability of the bandwidth with the external electric field
is a very distinctive feature of TDBG compared with TBG.

In conclusion, we have performed atomistic calculations for the atomic
and electronic structures of twisted double bilayer graphene.
In the equilibrium atomic structures, in-plane strain is present
mainly at the rotationally mismatched interface, while the out-of-plane
relaxation is significant in all of the four graphene layers.
Since the crystal structure lacks the inversion symmetry,
the electronic structure of TDBG can have an intrinsic band gap
at the charge neutral point over a large range of the twist angle.
At low twist angles, TDBG hosts well-isolated flat bands near the Fermi level,
which are generally narrower than the flat bands 
in TBG for $\theta > 1.1^\circ$.
Furthermore, we have shown that, by applying a vertical electric field,
the low-energy electronic structure of TDBG can be easily tuned,
and extremely narrow bands can be obtained in the electron side 
in a wide range of the twist angle, suggesting bigger chance of
correlated electronic states.
Our results provide basic electronic structures that serve
as a guide for exploring emergent correlated electron physics,
and illustrate the importance of the layer-number degree of freedom in 
tailoring flat-band physics in moir\'e superlattices.

\begin{acknowledgments}
  This work was supported by NRF of Korea (Grant No.~2011-0018306).
  Y.W.C. acknowledges support from NRF of Korea
  (Global Ph.D. Fellowship Program NRF-2017H1A2A1042152).
  Computational resources have been provided by KISTI
  Supercomputing Center (Project No.~KSC-2018-CRE-0097).
\end{acknowledgments}

\pagebreak
%%%%% SUPPLEMENTAL MATERIAL
\onecolumngrid
\setcounter{equation}{0}
\setcounter{figure}{0}
\setcounter{table}{0}
\renewcommand{\theequation}{S\arabic{equation}}
\renewcommand{\thefigure}{S\arabic{figure}}
\renewcommand{\bibnumfmt}[1]{[S#1]}
\renewcommand{\citenumfont}[1]{S#1}

\begin{center}

\textbf {
    Supplemental Material: Intrinsic Band Gap and Electrically Tunable Flat Bands in Twisted Double Bilayer Graphene}\\[.1cm]
Young Woo Choi and Hyoung Joon Choi$^*$\\[.0cm]
{\itshape Department of Physics, Yonsei University, Seoul 03722, Korea\\}
(Dated: \today)
\end{center}

\begin{center}
\setlength{\fboxrule}{0pt}

\fbox{\begin{minipage}{0.8\textwidth}
  \hspace{5pt} This supplemental material provides (i) comparison of electronic structures 
  of twisted bilayer graphene and twisted double bilayer graphene (TDBG) for 
  a wider energy range than in Fig.~2 in the main text, 
  (ii) effects of atomic-position relaxation on electronic structures of TDBG, and (iii) electronic 
  structures of nonrelaxed TDBG under electric fields.
  \end{minipage}}
\end{center}

\vspace{.0cm}
\twocolumngrid

\begin{figure}[b]
\includegraphics[width=8.7cm]{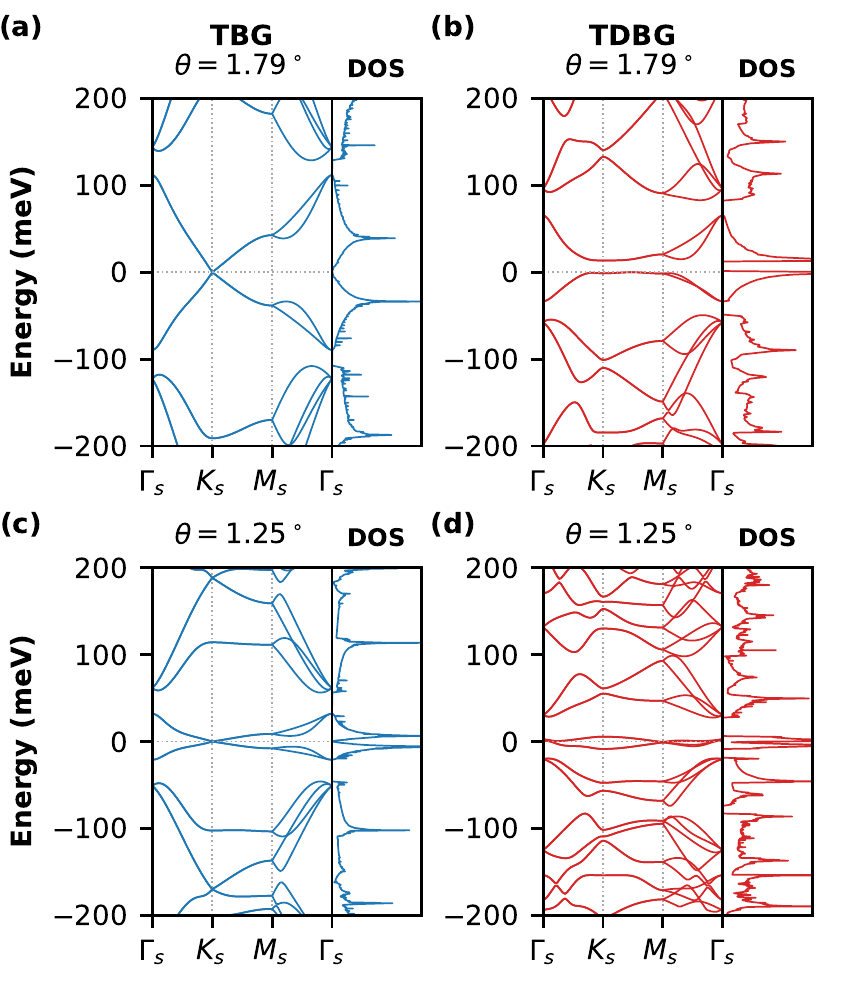}
\caption{\label{fig1}
  Electronic structures of (a,c) TBG and (b,d) TDBG with relaxed atomic positions 
  at (a,b) $\theta=1.79^\circ$ and (c,d) $\theta=1.25^\circ$.
  Electronic band structures and DOS are plotted for a wider energy range than in Fig.~2.
  In all cases, very narrow bands appear near zero energy and 
  band gaps are clearly shown above and below the narrow bands.
  }
\end{figure}

Figure~\ref{fig1} shows electronic structures of twisted bilayer graphene (TBG) and twisted double bilayer 
graphene (TDBG) with relaxed atomic positions at twisted angles of $\theta=1.79^\circ$ and $1.25^\circ$,
plotted for a wider energy range than in Fig.~2 in the main text.
Bandwidths of narrow bands near zero energy in TDBG are about half of those in TBG. 
In both TBG and TDBG, band gaps are clearly present above and below the narrow bands.
At $\theta=1.25^\circ$, the density of states (DOS) shows multiple strong peaks 
at high energies as well as near flat-band energies.

\begin{figure}[b]
\includegraphics[width=8.7cm]{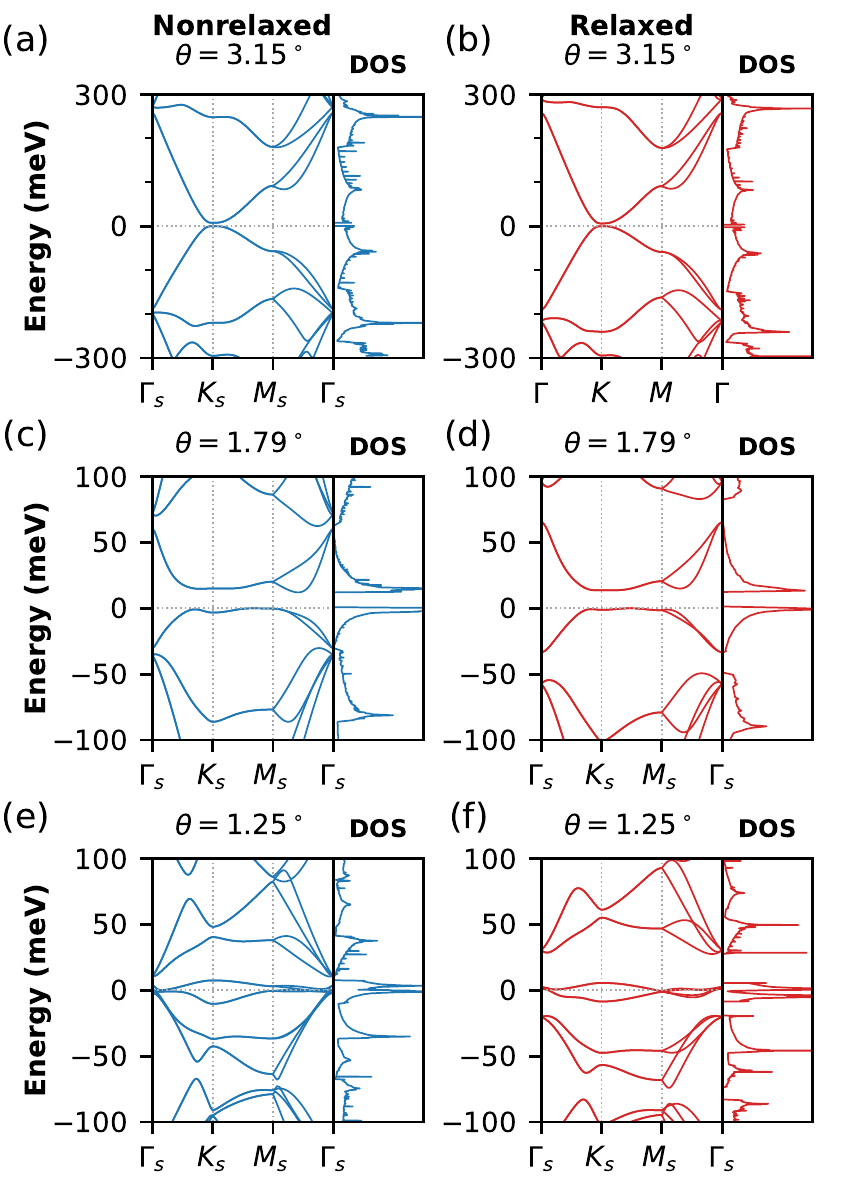}
\caption{\label{fig2}
  Band structures of TDBG with (a,c,e) nonrelaxed and (b,d,f) relaxed atomic positions
  at twist angles of (a,b) 3.15$^\circ$, (c,d) 1.79$^\circ$, and (e,f) 1.25$^\circ$.
  Effects of atomc-position relaxation become more important as the twist angle
  is lowered.
  }
\end{figure}

%\subsection{S2. Effects of lattice relaxation in TDBG}
Figure~\ref{fig2} shows electronic structures of TDBG at different twist angles 
before and after atomic-position relaxation.
Firstly, the band gap at the charge neutral point is present
both before and after the relaxation at $\theta$ = 3.15$^\circ$ and 1.79$^\circ$,
and the size of the band gap is almost unaffected by the relexation.
Bandwidths of narrow bands near zero energy are not affected significantly by the relaxation, either.
Secondly, below $\theta\sim2^\circ$, energy gaps open above and below the narrow bands 
after the relaxation, which become more pronounced as $\theta$ is lowered.

\begin{figure}[b]
\includegraphics[width=8.7cm]{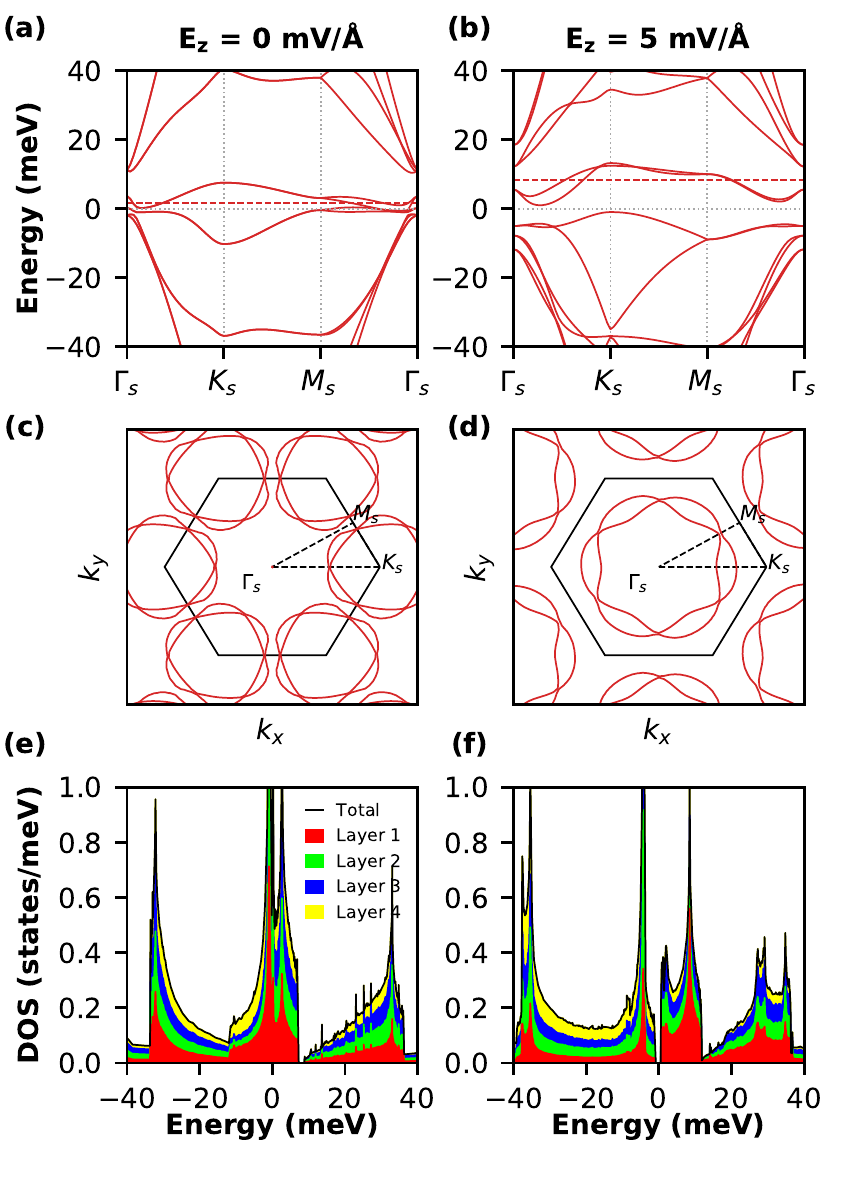}
\caption{\label{fig3}
    Electronic structures of TDBG with {\em nonrelaxed} atomic positions at $\theta=1.25^\circ$. 
    (a,b) Band structures with the vertical electric field $E_z$ of 
    (a) 0~mV/{\AA} and (b) 5~mV/{\AA}.
    Horizontal dashed lines denote half-filling energies of 
    electron-side flat bands.
    (c,d) Fermi surfaces at the half-filling energies of the electron-side
    flat bands with $E_z$ of (c) 0~mV/{\AA} and (d) 5~mV/{\AA}.
    (e,f) Electronic density of states with $E_z$ of (e) 0~mV/{\AA} 
    and (f) 5~mV/{\AA}.  Contributions of the 1st, 2nd, 3rd, and 4th 
    graphene layer to the total density of states are shown in red, green, 
    blue, and yellow, respectively.
    }
\end{figure}

Figure~\ref{fig3} shows electronic structures of nonrelaxed TDBG under electric fields.
Compared with the case of relaxed TDBG shown in Fig.~3 in the main text,
band dispersions of flat bands and Fermi surfaces at the half-filling energies
are generically similar.
However, without atomic-position relaxation, flat bands have wider bandwidths 
and they are not clearly separated from high energy bands.

%
% ****** End of file apssamp.tex ******

\end{document}